
%
\newbox\leftpage \newdimen\fullhsize \newdimen\hstitle \newdimen\hsbody
\tolerance=1000\hfuzz=2pt
\def\printertype{ps: }
\def\qms{\def\printertype{ln3: }
\ifx\answ\bigans\else\voffset=-.4truein\hoffset=.125truein\fi}
\def\bigans{b }
\message{ big or little (b/l)? }\read-1 to\answ
\ifx\answ\bigans\message{(This will come out unreduced.}
\magnification=1200\baselineskip=28pt plus 2pt minus 1pt
\hsbody=\hsize \hstitle=\hsize 
\else\message{(This will be reduced.} \let\lr=L
\magnification=1000\baselineskip=28pt plus 2pt minus 1pt
\voffset=-.31truein\vsize=7truein\hoffset=-.59truein
\hstitle=8truein\hsbody=4.75truein\fullhsize=10truein\hsize=\hsbody
\output={\ifnum\pageno=0 
  \shipout\vbox{\special{\printertype }\makeheadline
    \hbox to \fullhsize{\hfill\pagebody\hfill}}\advancepageno
  \else
  \almostshipout{\leftline{\vbox{\pagebody\makefootline}}}\advancepageno
  \fi}
\def\almostshipout#1{\if L\lr \count1=1 \message{[\the\count0.\the\count1]}
      \global\setbox\leftpage=#1 \global\let\lr=R
  \else \count1=2
    \shipout\vbox{\special{\printertype }
      \hbox to\fullhsize{\box\leftpage\hfil#1}}  \global\let\lr=L\fi}
\fi
%
\catcode`\@=11 
\newcount\yearltd\yearltd=\year\advance\yearltd by -1900

\def\Title#1#2{\nopagenumbers\abstractfont\hsize=\hstitle\rightline{#1}%
\vskip .5in\centerline{\titlefont #2}\abstractfont\vskip .5in\pageno=0}
\def\Date#1{\vfill\leftline{#1}\tenpoint\supereject\global\hsize=\hsbody%
\footline={\hss\tenrm\folio\hss}}
%

\def\draftmode{\message{ DRAFTMODE }\def\draftdate{{\rm preliminary draft:
\number\month/\number\day/\number\yearltd\ \ \hourmin}}%
\headline={\hfil\draftdate}\writelabels\baselineskip=20pt plus 2pt minus 2pt
 {\count255=\time\divide\count255 by 60 \xdef\hourmin{\number\count255}
  \multiply\count255 by-60\advance\count255 by\time
  \xdef\hourmin{\hourmin:\ifnum\count255<10 0\fi\the\count255}}}
\def\nolabels{\def\wrlabel##1{}\def\eqlabel##1{}\def\reflabel##1{}}
\def\writelabels{\def\wrlabel##1{\leavevmode\vadjust{\rlap{\smash%
{\line{{\escapechar=` \hfill\rlap{\sevenrm\hskip.03in\string##1}}}}}}}%
\def\eqlabel##1{{\escapechar-1\rlap{\sevenrm\hskip.05in\string##1}}}%
\def\reflabel##1{\noexpand\llap{\noexpand\sevenrm\string\string\string##1}}}
\nolabels
%
\global\newcount\secno \global\secno=0
\global\newcount\meqno \global\meqno=1
\def\newsec#1{\global\advance\secno by1\message{(\the\secno. #1)}
\global\subsecno=0\xdef\secsym{\the\secno.}\global\meqno=1
\bigbreak\bigskip\noindent{\bf\the\secno. #1}\writetoca{{\secsym} {#1}}
\par\nobreak\medskip\nobreak}
\xdef\secsym{}
\global\newcount\subsecno \global\subsecno=0
\def\subsec#1{\global\advance\subsecno by1\message{(\secsym\the\subsecno. #1)}
\ifnum\lastpenalty>9000\else\bigbreak\fi
\noindent{\it\secsym\the\subsecno. #1}\writetoca{\string\quad
{\secsym\the\subsecno.} {#1}}\par\nobreak\medskip\nobreak}
\def\appendix#1#2{\global\meqno=1\global\subsecno=0\xdef\secsym{\hbox{#1.}}
\bigbreak\bigskip\noindent{\bf Appendix #1. #2}\message{(#1. #2)}
\writetoca{Appendix {#1.} {#2}}\par\nobreak\medskip\nobreak}
%
%
\def\eqnn#1{\xdef #1{(\secsym\the\meqno)}\writedef{#1\leftbracket#1}%
\global\advance\meqno by1\wrlabel#1}
\def\eqna#1{\xdef #1##1{\hbox{$(\secsym\the\meqno##1)$}}
\writedef{#1\numbersign1\leftbracket#1{\numbersign1}}%
\global\advance\meqno by1\wrlabel{#1$\{\}$}}
\def\eqn#1#2{\xdef #1{(\secsym\the\meqno)}\writedef{#1\leftbracket#1}%
\global\advance\meqno by1$$#2\eqno#1\eqlabel#1$$}
%
\newskip\footskip\footskip14pt plus 1pt minus 1pt 
\def\f@@t{\baselineskip\footskip\bgroup\aftergroup\@foot\let\next}
\setbox\strutbox=\hbox{\vrule height9.5pt depth4.5pt width0pt}
\global\newcount\ftno \global\ftno=0
\def\foot{\global\advance\ftno by1\footnote{$^{\the\ftno}$}}
%
\newwrite\ftfile
\def\footend{\def\foot{\global\advance\ftno by1\chardef\wfile=\ftfile
$^{\the\ftno}$\ifnum\ftno=1\immediate\openout\ftfile=foots.tmp\fi%
\immediate\write\ftfile{\noexpand\smallskip%
\noexpand\item{f\the\ftno:\ }\pctsign}\findarg}%
\def\footatend{\vfill\eject\immediate\closeout\ftfile{\parindent=20pt
\centerline{\bf Footnotes}\nobreak\bigskip\input foots.tmp }}}
\def\footatend{}
%
%
\global\newcount\refno \global\refno=1
\newwrite\rfile
\def\ref{[\the\refno]\nref}
\def\nref#1{\xdef#1{[\the\refno]}\writedef{#1\leftbracket#1}%
\ifnum\refno=1\immediate\openout\rfile=refs.tmp\fi
\global\advance\refno by1\chardef\wfile=\rfile\immediate
\write\rfile{\noexpand\item{#1\ }\reflabel{#1\hskip.31in}\pctsign}\findarg}
\def\findarg#1#{\begingroup\obeylines\newlinechar=`\^^M\pass@rg}
{\obeylines\gdef\pass@rg#1{\writ@line\relax #1^^M\hbox{}^^M}%
\gdef\writ@line#1^^M{\expandafter\toks0\expandafter{\striprel@x #1}%
\edef\next{\the\toks0}\ifx\next\em@rk\let\next=\endgroup\else\ifx\next\empty%
\else\immediate\write\wfile{\the\toks0}\fi\let\next=\writ@line\fi\next\relax}}
\def\striprel@x#1{} \def\em@rk{\hbox{}}
\def\lref{\begingroup\obeylines\lr@f}
\def\lr@f#1#2{\gdef#1{\ref#1{#2}}\endgroup\unskip}

\def\addref#1{\immediate\write\rfile{\noexpand\item{}#1}} 
\def\startrefs#1{\immediate\openout\rfile=refs.tmp\refno=#1}
\def\xref{\expandafter\xr@f}\def\xr@f[#1]{#1}
\def\refs#1{[\r@fs #1{\hbox{}}]}
\def\r@fs#1{\vphantom{\hphantom{#1}}\edef\next{#1}\ifx\next\em@rk%
\def\next{}\else\ifx\next#1\xref #1\else#1\fi\let\next=\r@fs\fi\next}
%

%
\newwrite\ffile\global\newcount\figno \global\figno=1
\def\fig{fig.~\the\figno\nfig}
\def\nfig#1{\xdef#1{fig.~\the\figno}%
\writedef{#1\leftbracket fig.\noexpand~\the\figno}%
\ifnum\figno=1\immediate\openout\ffile=figs.tmp\fi\chardef\wfile=\ffile%
\immediate\write\ffile{\noexpand\medskip\noexpand\item{Fig.\ \the\figno. }
\reflabel{#1\hskip.55in}\pctsign}\global\advance\figno by1\findarg}
\def\xfig{\expandafter\xf@g}\def\xf@g fig.\penalty\@M\ {}
\def\figs#1{figs.~\f@gs #1{\hbox{}}}
\def\f@gs#1{\edef\next{#1}\ifx\next\em@rk\def\next{}\else
\ifx\next#1\xfig #1\else#1\fi\let\next=\f@gs\fi\next}
\newwrite\lfile
{\escapechar-1\xdef\pctsign{\string\%}\xdef\leftbracket{\string\{}
\xdef\rightbracket{\string\}}\xdef\numbersign{\string\#}}

\def\writestop{\def\writestoppt{\immediate\write\lfile{\string\pageno%
\the\pageno\string\startrefs\leftbracket\the\refno\rightbracket%
\string\def\string\secsym\leftbracket\secsym\rightbracket%
\string\secno\the\secno\string\meqno\the\meqno}\immediate\closeout\lfile}}
\def\writestoppt{}\def\writedef#1{}
\def\seclab#1{\xdef #1{\the\secno}\writedef{#1\leftbracket#1}\wrlabel{#1=#1}}
\def\subseclab#1{\xdef #1{\secsym\the\subsecno}%
\writedef{#1\leftbracket#1}\wrlabel{#1=#1}}
\newwrite\tfile \def\writetoca#1{}
\def\leaderfill{\leaders\hbox to 1em{\hss.\hss}\hfill}
\def\writetoc{\immediate\openout\tfile=toc.tmp
   \def\writetoca##1{{\edef\next{\write\tfile{\noindent ##1
   \string\leaderfill {\noexpand\number\pageno} \par}}\next}}}

%
\catcode`\@=12 
%
\ifx\answ\bigans
\font\titlerm=cmr10 scaled\magstep3 \font\titlerms=cmr7 scaled\magstep3
\font\titlermss=cmr5 scaled\magstep3 \font\titlei=cmmi10 scaled\magstep3
\font\titleis=cmmi7 scaled\magstep3 \font\titleiss=cmmi5 scaled\magstep3
\font\titlesy=cmsy10 scaled\magstep3 \font\titlesys=cmsy7 scaled\magstep3
\font\titlesyss=cmsy5 scaled\magstep3 \font\titleit=cmti10 scaled\magstep3
\else
\font\titlerm=cmr10 scaled\magstep4 \font\titlerms=cmr7 scaled\magstep4
\font\titlermss=cmr5 scaled\magstep4 \font\titlei=cmmi10 scaled\magstep4
\font\titleis=cmmi7 scaled\magstep4 \font\titleiss=cmmi5 scaled\magstep4
\font\titlesy=cmsy10 scaled\magstep4 \font\titlesys=cmsy7 scaled\magstep4
\font\titlesyss=cmsy5 scaled\magstep4 \font\titleit=cmti10 scaled\magstep4
\font\absrm=cmr10 scaled\magstep1 \font\absrms=cmr7 scaled\magstep1
\font\absrmss=cmr5 scaled\magstep1 \font\absi=cmmi10 scaled\magstep1
\font\absis=cmmi7 scaled\magstep1 \font\absiss=cmmi5 scaled\magstep1
\font\abssy=cmsy10 scaled\magstep1 \font\abssys=cmsy7 scaled\magstep1
\font\abssyss=cmsy5 scaled\magstep1 \font\absbf=cmbx10 scaled\magstep1
\skewchar\absi='177 \skewchar\absis='177 \skewchar\absiss='177
\skewchar\abssy='60 \skewchar\abssys='60 \skewchar\abssyss='60
\fi
\skewchar\titlei='177 \skewchar\titleis='177 \skewchar\titleiss='177
\skewchar\titlesy='60 \skewchar\titlesys='60 \skewchar\titlesyss='60
\def\titlefont{\def\rm{\fam0\titlerm}
\textfont0=\titlerm \scriptfont0=\titlerms \scriptscriptfont0=\titlermss
\textfont1=\titlei \scriptfont1=\titleis \scriptscriptfont1=\titleiss
\textfont2=\titlesy \scriptfont2=\titlesys \scriptscriptfont2=\titlesyss
\textfont\itfam=\titleit \def\it{\fam\itfam\titleit} \rm}
\ifx\answ\bigans\def\abstractfont{\tenpoint}\else
\def\abstractfont{\def\rm{\fam0\absrm}
\textfont0=\absrm \scriptfont0=\absrms \scriptscriptfont0=\absrmss
\textfont1=\absi \scriptfont1=\absis \scriptscriptfont1=\absiss
\textfont2=\abssy \scriptfont2=\abssys \scriptscriptfont2=\abssyss
\textfont\itfam=\bigit \def\it{\fam\itfam\bigit}
\textfont\bffam=\absbf \def\bf{\fam\bffam\absbf} \rm} \fi
\def\tenpoint{\def\rm{\fam0\tenrm}
\textfont0=\tenrm \scriptfont0=\sevenrm \scriptscriptfont0=\fiverm
\textfont1=\teni  \scriptfont1=\seveni  \scriptscriptfont1=\fivei
\textfont2=\tensy \scriptfont2=\sevensy \scriptscriptfont2=\fivesy
\textfont\itfam=\tenit \def\it{\fam\itfam\tenit}
\textfont\bffam=\tenbf \def\bf{\fam\bffam\tenbf} \rm}
%
%

\hyphenation{anom-aly anom-alies coun-ter-term coun-ter-terms}
\def\inv{^{\raise.15ex\hbox{${\scriptscriptstyle -}$}\kern-.05em 1}}

\def\Dsl{\,\raise.15ex\hbox{/}\mkern-13.5mu D} 
\def\dsl{\raise.15ex\hbox{/}\kern-.57em\partial}

\font\bigit=cmti10 scaled \magstep1
\def\lspace{\ifx\answ\bigans{}\else\qquad\fi}
\def\lbspace{\ifx\answ\bigans{}\else\hskip-.2in\fi} 
\def\boxeqn#1{\vcenter{\vbox{\hrule\hbox{\vrule\kern3pt\vbox{\kern3pt
        \hbox{${\displaystyle #1}$}\kern3pt}\kern3pt\vrule}\hrule}}}
\def\mbox#1#2{\vcenter{\hrule \hbox{\vrule height#2in
                \kern#1in \vrule} \hrule}}  
%

\def\darr#1{\raise1.5ex\hbox{$\leftrightarrow$}\mkern-16.5mu #1}

\def\half{{\textstyle{1\over2}}} 
\def\roughly#1{\raise.3ex\hbox{$#1$\kern-.75em\lower1ex\hbox{$\sim$}}}

\Title{\vbox{\baselineskip12pt\hbox{SU-HEP-4241-512}\hbox{TPI-MINN-92/35-T}}}
{{\vbox{\baselineskip12pt{\centerline{The Cosmological Kibble Mechanism in
the Laboratory:}
\vskip2pt \centerline{ String Formation in Liquid Crystals}}
}}}
\medskip
{\baselineskip12pt\centerline{ Mark J. ~Bowick\footnote{$^1$}
{ E-mail: Bowick@suhep.bitnet; Chandar@suhep.bitnet and Schiff@suhep.bitnet.},
L. ~Chandar$^1$ and Eric A.~Schiff$^1$}
\medskip
\centerline{ Physics Department }
\centerline{ Syracuse University}
\centerline{ Syracuse, NY 13244-1130, USA}}
\bigskip
{\baselineskip12pt\centerline{ Ajit M. Srivastava\footnote{$^2$}
{ E-mail: Ajit@umnacvx.bitnet ; address after 1st Sept.'92: Institute for
Theoretical Physics, University of California, Santa Barbara, CA 93106.}}
\centerline{ Theoretical Physics Institute }
\centerline{ University of Minnesota}
\centerline{ Minneapolis, MN 55455, USA}}
\bigskip
\centerline{\bf Abstract}
\medskip
\baselineskip28pt
We have observed the production of strings (disclination lines and loops) via
the Kibble mechanism of domain (bubble) formation in the isotropic to nematic
phase transition of a sample of uniaxial nematic liquid crystal. The probablity
of string formation per bubble is measured to be $0.33 \pm 0.01$. This is in
good agreement with the theoretical value $1/ \pi$ expected in two dimensions
for the order parameter space $S^2/{\bf Z}_2$ of a simple uniaxial nematic
liquid crystal.
\noindent
\Date{July 28, 1992}
\vfill \eject
%
%
\nref\deg{P.G. de Gennes, {\it The Physics of Liquid
Crystals} (Clarendon Press, Oxford, 1974).}
\nref\mer
{N.D. Mermin, Rev. Mod. Phys. {\bf 51} (1979) 591.}
\nref\bou
{Y. Bouligand, in {\it Les Houches Session XXXV}, {\it Physics of Defects},
R. Balian, Ed. (North Holland, Dordrecht, 1981).}
\nref\kle
{M. Kl\'eman, {\it Points, Lines and
Walls in Liquid Crystals, Magnetic Systems and Various
Ordered Media} (Wiley, Chichester, 1983).}
\nref\chra{S. Chandrasekhar and G.S. Ranganath, {\it The Structure
and Energetics of Defects in Liquid Crystals}, Adv. Phys. {\bf 35}
(1986) 507-596.}%
\nref\cdty{I. Chuang, R. Durrer, N. Turok and B. Yurke,
Science {\bf 251} (1991) 1336; B. Yurke, A.N. Pargellis, I. Chuang and N.
Turok, Princeton University preprint PUPT-91-1274 (1991).}%
\nref\cty{I. Chuang, N. Turok and B. Yurke,
Phys. Rev. Lett. {\bf 66} (1991) 2472.}%
Nematic Liquid Crystals (NLCs) form a class of matter with a rich variety of
topological defects. While the classification and static properties of these
defects have been thoroughly studied \refs{\deg{--}\chra} the mechanism of
their formation and their dynamics is relatively unexplored in the laboratory.
Recently the evolution in three dimensions of the dense tangle of string
defects (disclinations) produced by a pressure-induced isotropic to nematic
phase transition of a sample of NLC (``K15") was studied experimentally
\refs{\cdty,\cty}. It was shown that the string length density (total length in
string per volume) decays linearly in time and that the coarsening of the
string density obeys the scaling hypothesis $\rho(t) \approx {1 \over
\xi(t)^2}$ where the length scale $\xi(t)$ is assumed to be both the mean
radius of curvature of strings and the mean separation between strings and
grows in time as $t^{\nu}$ with $\nu = 0.51 \pm 0.02$. It was also shown that a
string loop of radius $R(t)$ shrinks in time as $R(t) \propto (t_0 -
t)^{\alpha}$, with the exponent $\alpha$ being $0.5 \pm 0.03$. This agrees with
the exponent $\half$ obtained from an analysis of the nematodynamic equations
of a disclination line moving through a nematic medium with a constant velocity
$v$ \refs{\deg,\cty}.

\nref\kib{T. Kibble, J. Phys. {\bf A9} (1976) 1387.}%
\nref\zur{W.H. Zurek, Nature {\bf 317} (1985) 505.}%
An important motivation for this recent work was to verify several of the
string interactions postulated in the study of cosmic strings and fundamental
strings.
In this letter we report our observations of the {\it formation} of
string defects in the supercooling of the NLC $K15$.  Preceeding
the production of strings we see the formation of uncorrelated domains
(bubbles) as described theoretically by Kibble \kib\ some years ago in his
pioneering analysis of the production of defects (domain walls, cosmic strings
and monopoles) in the cooling of the early Universe.
In our experiment disclinations are analogous to {\it global} cosmic strings
rather than {\it local} cosmic strings as there are no analogues of gauge
fields in simple uniaxial NLCs.
The idea of testing cosmological theories of string formation in
condensed matter systems was in fact proposed by Zurek \zur, who suggested a
cryogenic experiment to detect superfluid helium flow arising from the random
formation of uncorrelated domains produced by a rapid pressure quench in an
annular geometry.

NLCs are {\it thermotropic} materials typically consisting of rod-like
molecules roughly $20 \AA$ in length and $5 \AA$ in width. They often have two
benzene rings associated with different functions and short flexible exterior
chains. They exhibit a high-temperature {\it isotropic} phase characterized by
disorder with respect to translational and rotational symmetries and a
low-temperature {\it nematic} phase in which there is long-range orientational
order but translational disorder. The nematic phase is thus liquid with respect
to translations and crystalline with respect to orientations. Configurations of
NLCs are specified by a director field ${\bf n}(\vec{x})$. This is a
three-dimensional directionless unit vector giving at each point $\vec{x}$ the
local orientation of the rod molecules. An order parameter for the first-order
isotropic to nematic phase transition is $\langle {{3cos^2\theta - 1} \over 2}
\rangle$ where $\theta$ is the angle of the director field to some arbitrary
fixed axis. This order parameter vanishes in the isotropic phase and is
non-zero in the nematic phase.
For $K15$ the isotropic to nematic transition occurs at $T_{NI} =
35.3^{\circ}C$ at atmospheric pressure.
The easily accessible temperatures (typically $0^{\circ}-200^{\circ}C$)
of nematic phase transitions is one reason
these systems are simple to work with in the laboratory (especially
compared to superfluid helium).

The effective energy density describing NLCs is similar to that of an $O(3)$
non-linear
sigma model except that the NLC system is invariant only under simultaneous
spatial and internal rotations rather than independent spatial
and internal rotations \cdty. This means that the effective energy density for
NLCs consists of three terms of different symmetry associated with splay, twist
and bend deformations respectively \chra, as well as a surface term.
Corresponding to
each term is a potentially different elastic (Frank) constant.
The non-linear sigma model field theory is recovered only in the isotropic
limit
in which all three elastic constants are equal (the one-constant
approximation).

NLCs exhibit three classes of topological defects. The space of ground-state
nematic configurations
${\cal M}$ (ground-state manifold)
is given by $SO(3)/D_\infty$, where $SO(3)$ is the symmetry group of the
disordered (isotropic) phase and $D_\infty$ is the unbroken subgroup consisting
of rotations
about the molecular axis and $180^{\circ}$ rotations about axes perpendicular
to the
molecular axis. This space is isomorphic to $S^2/{\bf Z}_2$, the two-sphere
with
antipodal points identified, which in turn is isomorphic to the
projective plane ${\bf RP}^2$.
The existence of topological defects is determined by the homotopy groups of
the ground-state manifold.
The non-vanishing homotopy groups of ${\cal M}$ are $\pi_1({\cal M})={\bf
Z}_2$,
$\pi_2({\cal M})={\bf Z}$ and $\pi_3({\cal M})={\bf Z}$. Thus there are ${\bf
Z}_2$
disclinations (string or line defects), integrally-charged monopoles (point
defects) and integrally charged
{\it texture}
defects (whose field theory analogues are Skyrmeons) associated with
non-trivial elements of the first,
second and third homotopy groups respectively \mer.
In this letter we will be concerned only with the $\pi_1$ defects.
Note that non-trivial strings arise from the lack of polarity of liquid-crystal
molecules
which gives rise to the director field identification ${\bf n} \equiv {\bf
-n}$.
This is encoded in the ${\bf Z}_2$ isotropy group of ${\cal M} = S^2/{\bf
Z}_2$.
A path in ${\cal M}$ which winds from ${\bf n}$ to ${\bf -n}$ as one goes
around a
closed loop in space will not be continuously deformable to a point and
hence will represent a non-trivial disclination (string) defect.

To see more clearly the director field configurations associated with
disclinations we now give a very brief description of them.
The basic string defect along the $z$-axis is characterized by the planar
configuration
of the director field \chra\
\eqn\disc{    n_x = {\rm cos} \phi, \quad n_y = {\rm sin} \phi \quad {\rm
and}\quad n_z = 0,}
where we have taken the director to lie in the $x-y$ plane.
For an isotropic NLC (all elastic constants equal) the equations of motion
minimizing the free energy are
\eqn\eom{ \nabla^2 \phi = 0.}
Solutions independent of the radial coordinate $r=\sqrt{x^2 + y^2}$ are given
by
\eqn\sols{ \phi = s\alpha + c,}
where $\alpha = {\rm tan}^{-1}(y/x)$ and $c$ is a constant between $0$ and
$\pi$.
In the core of the string (of typical radius $100 \AA$)
the director changes discontinuously and the most
reasonable assumption is that the system is in the
isotropic phase \kle.
The parameter $s$ labels the strength of the disclination.
As the core of the
string is encircled the director rotates
by an angle $2 \pi s$.
The basic disclinations are the type-$\pm \half$ disclinations. They can be
continuously rotated
into each other and so are topologically equivalent. Integral strength
disclinations
are continuously deformable to trivial configurations without a singular core
by escape into the third dimension \mer\ and hence are topologically
trivial although their dynamics is still of considerable interest.
The energy density per unit length of a disclination is proportional to the
square of its strength and hence the strength $\pm \half$ disclinations
dominate the statistical mechanics of these systems.

The formation of topological defects
in a phase transition as described by Kibble \refs{\kib} in the context of
the early Universe, arises due to the existence of uncorrelated
regions of space (domains). As these domains
come into contact with each other, the variation of the order parameter
field from one domain to another may be such that topological defects get
trapped at the junctions of these domains. The probability of defect formation
can be calculated by assuming that the order parameter field varies
randomly from one domain to another. Let us illustrate this by
considering a simpler case of 2 spatial dimensions when the ground state
manifold is a circle S$^1$. The value of the order parameter field in
the ground state
is then given by an angle $\theta$ between 0 and 2$\pi$.  The strings will
arise when $\theta$ winds by 2$\pi$ around a closed path in space.
Consider now the situation when three domains meet at a point and further that
$\theta$ takes random values (say $\theta_1, \theta_2$ and $\theta_3$)
in these three domains. The value of $\theta$
in between any two domains will be governed by energetic considerations and
will be such that it amounts to smallest variation of $\theta$ from one
domain to another. With this, the string formation at the junction will
correspond to the case when one of the angles (say $\theta_3$) is such that
$\theta_1 + \pi < \theta_3 < \theta_2 + \pi$. [Assuming that $\theta_2 + \pi
> \theta_1 + \pi$, otherwise the order of $\theta_1$ and $\theta_2$ should
be reversed. All angles here are taken mod($2\pi$).]
 The maximum and minimum
values of this angular span available for $\theta_3$ for string formation
are $\pi$ and 0 with an average of $\pi/2$. As $\theta_3$ can assume any value
between 0 and 2$\pi$, the probability that a string will form at
the junction is 1/4. The calculation of probability of formation of
other defects (and for other ground state manifolds) can be carried out
along similar lines. For NLCs
the ground state manifold is ${\bf RP^2}$ and the probability of
string formation  in 2 spatial dimensions
can be calculated to be 1/$\pi$ (see
\nref\vac{T. Vachaspati, Phys. Rev. {\bf D44} (1991) 3723.}\vac\ ) .

 It is known that for NLCs, the isotropic-nematic phase transition is of
first order \nref\onsg{L. Onsager, Ann. N.Y. Acad. Sci. {\bf Vol. 51}
(1949) 627.}\onsg\ and proceeds through the proccess of bubble nucleation.
As a sample of  NLC supercools below $T_{NI}$ it becomes favorable
for domains (bubbles) of true
ground state (nematic phase) to  form in the
false ground state medium (isotropic phase). Very small bubbles have too
much surface energy and collapse again.
Bubbles beyond a certain size (the critical bubbles) gain enough bulk energy to
overcome the surface energy
and begin to grow (see \nref\ajit{A.M. Srivastava, Phys. Rev. {\bf D45}
(1992) R3304; Phys. Rev. {\bf D46} (1992), in press.}\ajit\
for a discussion of vortex formation by
bubbles). The director field will randomly take values in ${\cal M}$
in different bubbles and will be roughly uniform inside a given bubble
(for energetic reasons). As the bubbles collide, the actual correlation
length will depend on the bubble size as well as on the rate of bubble
collision and when bubble collisions are frequent, each bubble will
represent an uncorrelated domain as used in the Kibble mechanism. [The
correlation length here is similar to the horizon in the cosmological setting
in which different domains are causally disconnected and hence cannot
communicate with each other.] As the system cools the initially chaotic
variation of the director field will smoothen as various neighboring domains
coalesce and attempt to lower gradient energies by aligning their associated
director fields. As described above, however, there may be topological
obstructions to neighboring domains aligning completely. If there are
non-trivial paths in ${\cal M}$ traced out by neighboring domains then $\pi_1$
( string or disclination) defects are trapped in the system somewhat like flux
tubes in Type-II superconductors. This is the process we have observed in the
experiment described below. One of the key quantities of interest in this
mechanism of string production via bubbles is the probability of string
formation  per bubble. We have determined the typical domain (bubble) size and
net string length produced thereby experimentally determining the value of
the string formation probability.

For our observations we used the NLC $K15$
($4$-cyano-$4'$-n-pentylbiphenyl)
manufactured by BDH Chemicals, Ltd..
We employed
an Olympus, Inc. model BH phase contrast microscope with a 10X objective.  The
microscope was equipped with a monochrome television camera and a standard
videocassette recorder.

 We tried two different approaches to observing the
isotropic - nematic phase transition.  In the first a freely suspended film of
$K15$ was created by dipping a wire loop 3 mm in diameter in a sample of
$K15$. This wire loop was soldered to an $18\Omega$ resistor which acted as a
heater.  The loop and heater assembly were fastened to an aluminum slide with a
keyhole machined to accomodate them.  With this apparatus we first heated the
wire loop until the film entered the isotropic phase;  in this phase the
microscope image was completely structureless.  We then switched off the
electrical heater and allowed the film to cool. A similar approach
was used previously by Chuang, {\it et al} \cdty\ for the observation of
the evolution of string defects, which formed very rapidly as the film entered
its nematic phase.  Our results in this regime largely reproduced this previous
work.

This apparatus proved to be unsuitable for observing homogeneous bubble
formation and the subsequent generation of strings.  A circular front of
nematic phase formed at the boundary of the film and propagated through
the central region of the film before we could finish observing the evolution
of the homogeneously nucleated bubbles.  We tried to reduce this effect by
slowing down the cooling rate by slowly reducing the current through the
resistor and by shining an auxiliary illuminator during the measurements.
This improved the production of bubbles but did not eliminate the problem
of the circular front. We therefore tried observing the
phase transitions by using a different method by using
a thin drop of K15 placed on an untreated, clean microscope
slide.  For this arrangement heating was done using an illuminator;
 we were able to obtain satisfactory images of bubble formation and evolution.

\nfig\fone{A series of five images showing the isotropic to nematic phase
transition in a drop of K15 on an untreated microscope slide;  the images show
a width of 0.7 mm of the drop.  Note the stages of bubble nucleation and
growth, bubble coalescence and string formation, and string coarsening.  The
delay times for each image (referred to the first frame showing discernible
bubbles) are (a) 2 s, (b) 3 s, (c) 5 s, (d) 11 s, and (e) 23 s. Scale
of the pictures is given in (f).}

One set of such images is reproduced in Fig. 1.  Fig. 1(a) shows the numerous
small isolated bubbles of nematic phase which form first.  Fig. 1(b) and 1(c)
show the images at short intervals later, as the nematic bubbles increase in
size. This occurs both by natural growth and also by coalescence. In the next
stage the organization of the NLC into bubbles is replaced by an image of a
homogeneous medium with entangled strings (Fig. 1(d)), which further evolve by
straightening, shrinking, and excising small loops of closed string (Fig.
1(e)).  String dynamics have been very well described previously
\refs{\cdty,\cty}.  The film becomes increasingly homogeneous in appearance on
longer time scales, as the string pattern coarsens.

We analyzed series of images such as Fig. 1 to estimate the string to bubble
ratio. In counting the number of bubbles we have to decide
which bubbles are relevant for string formation. For example between Fig. 1(b)
and 1(c) small bubbles seem to just get absorbed by bigger bubbles and we do
not see them contributing to any string formation.  This observation is
consistent with the results of the numerical simulation in \refs{\ajit} where
the collision of two large (critical) bubbles with one small (subcritical)
bubble was considered and it was found that even if a string forms in such a
collision, it gets pushed out of the smaller bubble due to large momentum of
the fast expanding walls of the large bubbles. Even in the collision
where all the bubbles involved are very small, we have not observed any
string formation. This could again be because a string
will get trapped at the collision point only when the bubble walls have
sufficient momentum which may not happen if bubbles are very small.
Even if some strings do form in the collisions of very tiny bubbles, these
strings will be very small (for isolated collisions) and will dissappear
quickly by shrinking. As we count the strings at a little later stage anyway
(when we can clearly distinguish them), these small strings would have
dissappeared by that time and hence no error will be caused by the
neglect of very small bubbles.

Guided by these considerations, we made a judgement of a
typical size for the bubble which seems to be large enough to
contribute to string formation and then consider coalescence and
growth of such bubbles. Bubbles $r$ times the average bubble size are assumed
to have arisen from the coalescence of $r$ average bubbles (as can be verified
by looking at the earlier stages of the bubble formation) and are thus counted
$r$ times.

The next step is to look at a photograph showing
these $N$ bubbles which will eventually
expand to fill up the entire region of the photograph
(of area $A$). As the bubbles will coalesce they will
not remain spherical but will actually squeeze into gaps.
The final image  will look roughly like a rectangular
region filled by $N$ squares (with typical side of a square $d$ roughly
equal to $\sqrt{A/N})$.
We estimated a count of about 55 ``effective" bubbles in Fig. 1(c);
the typical length scale was about 80 $\mu$m.
%
%
Our estimate of the probability that each of these bubbles will lead to a
string upon final coalescence was made as follows. Imagine that the
rectangular image above is a lattice of area $A$ with $m$ rows and $n$
columns such that $N=mn$. As strings form because of mismatches of the
director at bubble interfaces the maximum possible string length $L_{max}$ is
given by the total sum of the lengths of all the links on the lattice which is
$2(NA)^{1/2}$.  For Fig. 1(c) we estimate $L_{max} = 8.7 \rm\ mm$.  We then
examine a photograph of the earliest possible stage when one can discern
strings clearly and measure the total length of string $L_s$;  in Fig. 1(d) we
obtained 2.7 mm.  The probability of string formation is then taken to be
$P=L_s/L_{max} = 0.31$.

We repeated this analysis on several sequences, obtaining $P=0.33 \pm 0.01$
for the string formed per bubble.  Systematic errors are not negligible;  we
believe that this measurement provides a lower bound on the true value of $P$
for two reasons. First the total string length will have decreased from its
initial value by the time we are able to clearly resolve and measure $L_s$.
Secondly strings which wander significantly in the direction normal to the
plane of focus (but are still formed due to top bubble layers which we observe)
will have their length underestimated. The actual
underestimation may not be too bad since string wiggling will be dampened in
the time between their formation and our measurement of their length. Further
there is a selection effect in the experiment since we are able to see strings
clearly only when they do not deviate significantly from the plane of focus.
Thus we are effectively counting planar strings. This will suppress the
effects of the three-dimensionality of the bubble layers. In other words, the
bubbles we see are only the top few layers; similarly we see the strings at
the time they are roughly confined to these top layers. The numbers we obtain,
therefore, correspond to an effectively two-dimensional situation. Note that
for strictly 2 dimensional situation, the strings will form perpendicular to
the plane formed by the bubble layer. In the present case, however, strings
will bend around in the plane as they come towards the top (where the director
will eventually become normal to the surface this being the preferred boundary
condition for NLC-air interface) as well as when they squeeze through the
bubble layers underneath them.

We note here that the string to bubble ratio we have measured is in good
agreement with the value $1/ \pi \approx 0.32$ obtained by Vachaspati
\refs{\vac} for the ground-state manifold $S^2/{\bf Z}_2$ in {\it two}
dimensions, especially considering the above arguments that our estimates for
$P$ correspond to an effectively 2 dimensional situation.

We now turn to another interesting phenomenon we are able to observe in this
experiment.  We first cool our sample through the nematic phase transition so
that disclinations are produced and then superheat from the nematic phase to
the isotropic phase. Spherical isotropic bubbles are produced in the nematic
medium. These have a somewhat different more transparent appearance than the
nematic bubbles in an isotropic medium because of the different optical
properties of the two phases. In this way it is quite easy to distinguish the
different types of bubble. We observe very clearly in this process the
heterogeneous nucleation of isotropic bubbles along disclination lines like
beads on a necklace, see Fig. 2. This nucleation process is of great
technological and theoretical interest
\nref\gsms{J.D. Gunton, M.San Miguel and P.S. Sahni, in {\it Phase
Transitions and Critical Phenomenon}, Eds. C. Domb and J.L. Lebowitz
(Academic Press, New York, 1983), Vol.{\bf 8}, pp. 269-482.}%
\nref\mokl{L. Monette and W. Klein,
Phys. Rev. Lett. {\bf 68} (1992) 2336.}%
\refs{\gsms,\mokl}.
We feel that NLCs are a good testing ground to check experimentally
many of the theoretical predictions concerning the mechanism of the
nucleation process such as the size distribution of bubbles.

\nfig\fone{This series of three pictures shows the
heterogeneous nucleation of isotropic bubbles along disclination lines.
The delay times for each image, referred to the picture in (a), are
(b) 2 s, and (c) 4 s. Scale is given in (d).}

Finally we point out that the implications of the work described here extend
beyond the dynamics of the formation of disclinations in nematic liquid
cystals because of the close analogies with the formation of other
one-dimensional topological defects such as screw dislocations in crystals and
vortices in superfluid helium \refs{\chra,\zur}.

We thank Bernard Yurke for his valuable advice during his visit to Syracuse
University.  We would also like to express very warm thanks to Edward Lipson
and his biophysics group at Syracuse University for their generosity, both in
sharing their microscope apparatus and in instructing us in its use. Without
their support the research reported here would not have been possible. AMS
would like to thank Mark Hindmarsh for useful discussions on theoretical
estimates of string formation probability. The research of MJB was supported
by the Outstanding Junior Investigator Grant DOE DE-FG02-85ER40231. The
research of LC was supported by a Syracuse University Fellowship. The research
of AMS was supported by the Theoretical Physics Institute at the University of
Minnesota and by the US Department of Energy under contract number
DE-AC02-83ER40105. MJB acknowledges the hospitality of the theoretical physics
group at the Universit\`a di Roma, ``Tor Vergata" where this manuscript was
completed.
\footatend\vfill\supereject\immediate\closeout\rfile\writestoppt
\baselineskip=28pt\centerline{{\bf References}}\bigskip{\frenchspacing%
\parindent=20pt\escapechar=` \input refs.tmp\vfill\eject}\nonfrenchspacing
\vfill\eject\immediate\closeout\ffile{\parindent40pt
\baselineskip28pt\centerline{{\bf Figure Captions}}\nobreak\medskip
\escapechar=` \input figs.tmp\vfill\eject}
\bye